# The NAD(P)H:Flavin Oxidoreductase from *Escherichia coli* :

# Evidence for a New Mode of Binding for Reduced Pyridine Nucleotides

Vincent Nivière[‡]*, Franck Fieschi[§], Jean-Luc Décout[‡] and Marc Fontecave[‡]

[‡] Laboratoire de Chimie et Biochimie des Centres Rédox Biologiques, DBMS-CEA/ CNRS/ Université J. Fourier, Batiment K, 17 Avenue des Martyrs, 38054 Grenoble, Cedex 9, France

[§] Institut de Biologie Structurale, CEA/ CNRS/ Université J. Fourier, 41 Avenue des Martyrs, 38027 Grenoble, Cedex 1, France

Running Title : Substrate specificity of the NAD(P)H binding site of Fre

* To whom correspondence should be addressed.

Telephone : 33-(0)4-76-88-91-09. Fax : 33-(0)4-76-88-91-24.

E-mail:niviere@cbcrb.ceng.cea.fr

# SUMMARY

The NAD(P)H:flavin oxidoreductase from *Escherichia coli*, named Fre, is a monomer of 26.2 kDa that catalyzes the reduction of free flavins using NADPH or NADH as electron donor. The enzyme does not contain any prosthetic group but accommodates both the reduced pyridine nucleotide and the flavin in a ternary complex prior to oxidoreduction. The specificity of the flavin reductase for the pyridine nucleotide was studied by steady-state kinetics using a variety of NADP analogs. Both the nicotinamide ring and the adenosine part of the substrate molecule have been found to be important for binding to the polypeptide chain. However, in the case of NADPH, the 2'-phosphate group destabilized almost completely the interaction with the adenosine moiety. Moreover, NADPH and NMNH are very good substrates for the flavin reductase and we have shown that both these molecules bind to the enzyme almost exclusively by the nicotinamide ring. This provides evidence that the flavin reductase exhibits a unique mode for recognition of the reduced pyridine nucleotide. In addition, we have shown that the flavin reductase selectively transfers the pro-R hydrogen from the C-4 position of the nicotinamide ring and is therefore classified as an A side specific enzyme.

INTRODUCTION

Flavin reductases are enzymes defined by their ability to catalyze the reduction of free flavins (riboflavin, FMN or FAD) by using reduced pyridine nucleotides, NADPH or NADH (1). The products of these enzyme activities, protein-free reduced flavins, were suggested to have important biological functions as electron transfer mediators, even though the real physiological significance of these mediators has so far not been fully appreciated. In fact, *in vitro*, free reduced flavins can reduce ferric complexes or iron proteins very efficiently and it has been suggested that flavin reductases could play a key role in : iron metabolism (1, 2), activation of ribonucleotide reductase (3, 4) and reduction of methemoglobin (5, 6). There is also indirect evidence for their function in bioluminescence (7, 8) and oxygen reduction (9). Recently, flavin reductases have been found to be associated with oxygenases involved in the desulfurization process of fossil fuels (10) and antibiotic biosynthesis (11-13).

Up to now, two major classes of flavin reductases have been characterized. Class I enzymes do not contain any flavin prosthetic group and cannot be defined as flavoproteins (3, 14), whereas class II enzymes are canonical flavoproteins (15-17). These two classes also exhibit different enzymatic mechanisms for flavin reduction. Class I enzymes use a sequential mechanism (14), whereas class II use a Ping-Pong mechanism, as a consequence of having a redox cofactor bound to the polypeptide chain (17, 18).

The prototype for class I flavin reductases is an enzyme, named Fre [1], which was initially discovered in *Escherichia coli* as a component of a multienzymatic system involved in the activation of ribonucleotide reductase, a key enzyme in DNA biosynthesis (3, 4). Fre consists of a single polypeptide chain of 233 amino acids, with a molecular mass of 26,212 Da (19). It uses an electron donor which can be either NADPH or NADH and riboflavin (Rf) is the best substrate (3, 14). Steady-state kinetic studies have shown that Fre functions through an ordered mechanism, with NADPH binding first and riboflavin being the second substrate (14). Product order release has also been determined, with reduced flavin being the first product to be released and NADP$^+$ the second (14).

Although Fre is not a flavoprotein, several lines of evidence suggested that Fre belongs to a large family of flavoenzymes of which spinach ferredoxin-NADP$^+$ reductase (FNR) is the structural prototype (20). In spite of very low global sequence similarity, Fre contains a sequence motif of four amino acids, $R_{46}PFS_{49}$, similar to the $R_{93}LYS_{96}$ region of spinach FNR. In the FNR family, these conserved residues are involved in the interaction between the protein and the flavin isoalloxazine ring (20). In spinach FNR, Ser96 has been proposed to make a hydrogen bond to the N(5) position of the FAD cofactor (20) and was found to be essential for activity (21). Site-directed mutagenesis experiments allowed us to demonstrate that Ser49 of Fre (corresponding to Ser96 of spinach FNR) is indeed involved in flavin binding, and that it is also essential for activity (22). These results were the first to support the structural relationship between Fre and the members of the FNR family.

This was further confirmed by characterization of reaction intermediates which are formed during the catalytic cycle of Fre (23). For the flavoenzymes of the FNR family, rapid kinetic studies have demonstrated the occurrence of two distinct intermediates species during the catalytic cycle, identified as charge transfer complexes between pyridine nucleotide and the flavin cofactor (24-26). Similarly, in the case of Fre, the same two reaction intermediates of the catalytic cycle have been characterized (23). Together, these results now extend the structural similarity between Fre and the proteins of the FNR family to a functional level.

Substrate specificity studies on Fre using flavin analogs have shown that the isoalloxazine ring of the flavin is the only part of the molecule which is recognized by Fre (14). Such a weak interaction of Fre with the flavin provides a rational explanation why a protein related to a flavoprotein family can use a flavin as a substrate rather than as a cofactor (14).

In the present paper, using a variety of NAD(P) analogs, substrates or inhibitors, we have now studied the specificity of Fre for the reduced pyridine nucleotide, the other substrate. We have shown that Fre exhibits a unique mode of binding with the reduced pyridine nucleotide, which involves both the adenosine and the nicotinamide ring moiety of the molecule. However, when NADPH is used as electron donor, the presence of the 2'-phosphate group of the molecule destabilized almost completely the interaction with the adenosine moiety. NADPH, as NMNH, binds to Fre with the nicotinamide ring almost exclusively. In addition, we have shown that Fre selectively transfers the pro-R hydrogen from the C-4 position of the nicotinamide group and is therefore classified as an A-side specific enzyme.

EXPERIMENTAL PROCEDURES

*Biochemical and chemical reagents*

Riboflavin (Rf), FAD, lumichrome, NAD(P)H analogs, lyophilized yeast alcohol dehydrogenase, lyophilized *Thermoanaerobium brockii* alcohol dehydrogenase, 2-propanol-$d_8$ were obtained from Sigma. $^2H_2O$ (99.9+ atom % $^2H$) was from SDS. Others reagent-grade chemicals were obtained from either Sigma, Aldrich, or Bio-rad.

*Production of recombinant flavin reductase*

Purification of the recombinant *Escherichia coli* Fre was carried out using a two-step protocol, phenyl-Sepharose and Superdex 75 as previously reported (14). The protein concentration was determined using the Bio-Rad Protein Assay reagent (27) and bovine serum albumin as a standard protein.

*Enzyme assays*

Flavin reductase activity was determined from the decrease of the absorbance due to the oxidation of the reduced pyridine nucleotide in the presence of Rf or FAD, using a Varian Cary 1 Bio spectrophotometer. Activities were measured at 25 °C in 50 mM Tris-Cl buffer pH 7.6. Assays were performed in 1 cm path length cuvette (final volume 1 ml). When high concentrations of NAD(P)H or NMNH were investigated, a 0.1 cm path length cuvette was used (final volume 0.3 ml). When lumichrome was present in the enzymatic assays, the cuvette contained in addition 87 mM $Me_2SO$ final concentration as reported in (14). The reaction was initiated by adding 0.2-15 µg of enzyme, depending of the reduced pyridine nucleotide used. Activities with NADPH and NADH were followed at 340 nm ($\varepsilon$=6.22 mM$^{-1}$ cm$^{-1}$), with NMNH at 338 nm ($\varepsilon$=5.72 mM$^{-1}$ cm$^{-1}$), with thio-NADH at 396 nm ($\varepsilon$=11.3 mM$^{-1}$ cm$^{-1}$), with 3-acetyl NADH at 364 nm ($\varepsilon$=5.6 mM$^{-1}$ cm$^{-1}$). Enzyme activities were determined from the linear part of the progress curve, with less than 10% of reduced pyridine nucleotide utilized over the time course of the reaction.

*Kinetic analysis*

Reciprocal initial velocities were plotted against reciprocal substrate concentrations, and all plots were verified to be linear. Kinetic parameters ($v_m$, $K_m$, $K_{mapp}$) were determined from saturation curves, fitted with the equation $v_i=(v_m[S])/(K_m+[S])$, using a Levenberg-Marquardt algorithm. Inhibition constants ($K_i$) were determined using a replot of the $K_m$ apparent values ($K_{mapp}$) versus inhibitor concentration [I] and fitted with a straight line, assuming the equation for a competitive inhibitor : $K_{mapp}=(K_m/K_i)[I]+K_m$. When applicable, values are shown ± standard deviation.

*Substrate- product concentrations*

NAD(P) analogs and flavin concentrations were determined spectroscopically using the following extinction coefficients : adenosine ($\varepsilon_{260nm}$=14.9 mM$^{-1}$ cm$^{-1}$), AMP, ADP-ribose, ATP-ribose and 2', 5' ADP ($\varepsilon_{259nm}$=15.4 mM$^{-1}$ cm$^{-1}$), ß-NAD(P)$^+$ ($\varepsilon_{259nm}$=17.8 mM$^{-1}$ cm$^{-1}$), ß-NMN$^+$

($\varepsilon_{266nm}$=4.6 mM$^{-1}$ cm$^{-1}$), nicotinamide ($\varepsilon_{260nm}$=2.9 mM$^{-1}$ cm$^{-1}$), Rf ($\varepsilon_{450nm}$=12.5 mM$^{-1}$ cm$^{-1}$), FAD ($\varepsilon_{450nm}$= 11.3 mM$^{-1}$ cm$^{1}$), lumichrome ($\varepsilon_{356nm}$= 6.0 mM$^{-1}$ cm$^{-1}$).

*Determination of the stereospecificity of the hydride transfer*

[4R-$^2$H] NADH and [4R-$^2$H] NADPH were prepared from enzymatic reduction of NAD$^+$ and NADP$^+$ by 2-propanol-$d$8 as previously described (28, 29). As a control NADH and NADPH were prepared by using non deuterated isopropanol. $^1$H-NMR spectra were recorded at 25°C on a Varian U$^+$ 500 apparatus. The reaction mixture contained 3 mM [4R-$^2$H] NADH or [4R-$^2$H] NADPH in 1 ml of 50 mM of potassium phosphate buffer p$^2$H 8, 25 µg of Fre, 15 µM of Rf and was incubated at 25°C under aerobic conditions. The oxidation of [4R-$^2$H] NAD(P)H was monitored spectrophotometrically at 340 nm. After total oxidation, $^1$H-NMR spectra of the samples were recorded. The stereospecificity of NAD(P)H was calculated from the ratio [peak area of C4-H (δ 8.74 ppm)] / [average peak area of C2-H (δ 9.24 ppm) and C6-H (δ 9.05 ppm)].

RESULTS

*NAD(P)H analogs as substrates*

Scheme 1 shows the structures of the various NAD(P)H analogs studied. Table 1 shows the kinetic and thermodynamic parameters obtained with Fre in the presence of various NAD(P)H analogs as electron donors and riboflavin (Rf) as the electron acceptor. $K_d$ values for the different NAD(P)H analogs were calculated from the Dalziel mathematical treatment of the initial velocities data (30), which proved to be well-adapted in the case of Fre (14, 22). For a sequential bireactant mechanism, where A and B are the first and second substrate of the reaction respectively, the initial reaction velocity, $v_i$, can be expressed in a reciprocal form in terms of $\phi$'s parameters as suggested by Dalziel (30) :

$$e/v_i = \phi_o + \phi_A/[A] + \phi_B/[B] + \phi_{AB}/[A][B] \qquad (1)$$

where e is the enzyme concentration. The different $\phi$ parameters in equation 1 represent a combination of rate constants of each individual step of the sequential bireactant mechanism (30). A plot of the reciprocal of the concentration of A versus $e/v_i$, at a constant concentration of B, will have a slope equal to $\phi_A + \phi_{AB}/[B]$ and an Y-axis intercept equal to $\phi_o + \phi_B/[B]$. Secondary replots of the slopes and intercepts versus the reciprocal of the concentration of

B will give the values of the different $\phi$ parameters, as illustrated in Fig. 1. The $\phi$ parameters are related to the $K_m$, $k_{cat}$ and $K_d$ values of the reaction as follows :

$$K_m(A) = \phi_A/\phi_o; K_m(B) = \phi_B/\phi_o; K_d(A) = \phi_{AB}/\phi_B; k_{cat} = 1/\phi_o$$

$K_m$ and $k_{cat}$ values for the different reduced pyridine nucleotides obtained by Dalziel treatment are checked to be identical to $K_m$ and $k_{cat}$ values determined from steady-state kinetic analysis of the reaction during experiments in which NAD(P)H analogs are varied in the presence of saturating concentration of Rf. In all cases, the $K_m$ value for Rf determined in the presence of different reduced pyridine nucleotides remained in the 1-10 µM range (Table 1).

Unexpectedly, NMNH was found to be a very good substrate for the enzyme reduction of Rf. In Fig. 1, Fre activity was determined as a function of NMNH concentration at several levels of Rf. Double reciprocal plots show a series of intersecting lines to the left of the vertical axis, consistent with a sequential mechanism as observed in the case of NADPH, with Rf as the electron acceptor (14). The $k_{cat}$ value was significantly larger than that obtained with NADPH as the electron donor (Table 1). $K_m$ value for NMNH was found to be 3.5 and 14 times more elevated than the corresponding values for NADPH and NADH respectively and as a consequence the catalytic efficiency for NMNH, $k_{cat}/K_m$, was only reduced about 2.5 and 5-fold compared to that determined for NADPH and NADH respectively (Table 1). The $K_d$ value for NMNH was 116±17 µM, a value slightly larger than the $K_d$ value from NADPH

(54±9 µM), determined under the same experimental conditions (Table 1). These data thus strongly suggested that the adenosine phosphate part of the reduced pyridine nucleotide does not play a major role in the interaction with the Fre polypeptide chain and in the formation of the catalytic competent ternary complex with Rf (14, 23).

We have examined the inhibition pattern of $NMN^+$ when NMNH was used as the electron donor and Rf as the electron acceptor. When NMNH was varied in the presence of a fixed non-saturating concentration of Rf, inhibition by $NMN^+$ was found to be competitive with respect to NMNH (data not shown). A $K_i$ value for $NMN^+$ of 18 mM has been estimated (Table 2). These results strongly suggest that the first product to be released is the reduced Rf, followed by $NMN^+$, as with NADPH as electron donor (14).

Two NADH analogs, 3-acetyl NADH and thio-NADH, with limited modifications at the carboxamide group of the nicotinamide ring, were tested and also found to be substrates for the reaction catalyzed by Fre. However, they exhibited significantly lower $k_{cat}$ values compared to that of NADH (Table 1), which may reflect their higher redox potential values. Fre activity was determined as a function of 3-acetyl NADH concentration at several levels of Rf (data not shown). The double reciprocal plots of the initial velocities show a series of intersecting lines to the left of the vertical axis (data not shown), consistent with a sequential mechanism. $K_d$ value determined for 3-acetyl NADH was found to be about 6 times larger than the corresponding value for

NADPH, whereas its $K_m$ value was 13 and 52 times larger than the corresponding $K_m$ values for NADPH and NADH respectively (Table 1).

$K_m$ value for thio-NADH was found to be very low, 0.5 µM (Table 1), suggesting an increased affinity of Fre for this compound. Because of the low $K_m$ and $k_{cat}$ values, $K_d$ value for thio-NADH could not be accurately determined.

*Enzymatic mechanism with NADH as a substrate*

Fig. 2 shows Fre activity determined as a function of NADH concentration at several levels of Rf. Surprisingly, double reciprocal plots of the initial velocity versus NADH concentration showed a series of parallel lines. When Fre activity was determined as a function of Rf at several levels of NADH, double reciprocal plots showed again a series of parallel lines (data not shown). When plotting the initial rate data in the form $[S]/v_i$ versus $[S]$, where $[S]$ is the concentration of the varied substrate, we obtained a series of lines that all intersect on the $[S]/v_i$ axis (data not shown). This confirms the parallel line pattern of the double reciprocal plots (31). Such a pattern would suggest a Ping-Pong mechanism. In any case, data from Fig. 2 cannot be used to determine a $K_d$ value for NADH with the Dalziel treatment.

That the reduction of Rf by NADH could involve a Ping-Pong mechanism was intriguing, considering that Fre does not contain a redox cofactor and that NADPH was clearly shown to function with a sequential mechanism. To further investigate the possibility of a Ping-Pong mechanism, we carried out a series of inhibition studies, with NADH as the electron donor. Table 2 shows that AMP was found to be a competitive inhibitor with respect to NADH (with a $K_i$ value of 420±70 μM) and non-competitive inhibition was observed with respect to Rf. Lumichrome is a strong competitive inhibitor with respect to Rf, with a $K_i$ value of 60±9 nM, and an uncompetitive inhibitor with respect to NADH. These data do not support a Ping-Pong mechanism but are consistent with a sequential ordered mechanism with NADH binding first (32), as it has been previously reported for reduction of Rf by NADPH (14).

The parallel line pattern observed in Fig. 2 could then be explained by a dissociation constant value of NADH ($\phi_{AB}/\phi_B$) much smaller than its Michaelis constant ($\phi_A/\phi_o$), as suggested already by several authors (30, 31, 33). If this is the case, this makes $\phi_{AB}$ difficult to estimate and equation 1 would become : $e/v_i = \phi_o + \phi_A/[A] + \phi_B/[B]$, assuming a parallel lines pattern for a sequential ordered mechanism.

The kinetic mechanism of product release for the Fre catalyzed reaction using NADH and Rf as substrates has been determined by studying product inhibition. When NADH concentration was varied with a fixed, non saturating concentration of Rf, inhibition by $NAD^+$ was found to be mixed-type with respect to NADH (Fig. 3A, Table 2). Furthermore, when Rf concentration was varied with a fixed, non saturating concentration of NADH, $NAD^+$ was found to be mixed-type with respect to Rf (Fig. 3B, Table 2). These data strongly suggest that $NAD^+$ is the first product to be released (33).

In conclusion, the enzyme works with NADH almost as with NADPH through a sequential ordered mechanism and the parallel line pattern observed in Fig. 2 can be explained by NADH having a $K_d$ value much smaller than its $K_m$ value (8±1 µM, Table 1). However, the reaction with NADH differs from that with NADPH in that $NAD^+$ is the first reaction product to be released.

*NAD(P) analogs as dead-end inhibitors*

The data reported in Table 1 suggested that the nicotinamide ribose phosphate part of the reduced pyridine nucleotide played a major role in binding to Fre. We have tested a variety of NAD(P) analogs, all lacking the nicotinamide ring (Scheme 1), as dead-end enzyme inhibitors (Table 3). All NAD(P) analogs tested in Table 3 were found to be competitive inhibitors with respect to NADPH. In the case of a sequential ordered mechanism, the inhibition constant $K_i$ for a competitive inhibitor (I) with respect to the first substrate reflects the dissociation constant of the enzyme-inhibitor complex (EI) unless the second substrate can complex with EI to form an inactive ternary complex (34). In such a case, $K_i$ must be smaller than the true dissociation constant of the enzyme-inhibitor complex (34). In our case, an inhibition effect associated with an increased concentration of the second substrate, Rf, would be observed. This has been tested using AMP as the dead-end inhibitor. Fre activity was determined as a function of Rf concentration (varied in a 0.3-100 $K_m$ range) at a constant non saturating concentration of NADPH (100 µM) with and without 2 mM AMP. Double reciprocal plots of the initial velocities versus Rf concentration gave no evidence of a substrate-inhibition pattern (data not shown). We thus conclude that, under these conditions, no ternary Fre-Rf-AMP complex was formed and as a consequence the $K_i$ value reported in Table 3 can be interpreted as a true dissociation constant for the AMP-enzyme complex. By extension, we consider that all $K_i$ values reported in Table 3 are also true dissociation constants.

As shown in Table 3, $K_i$ values for 2', 5' ADP and ATP-ribose are much larger than the $K_d$ value for NADPH (3.7±1.0, 3.8±0.8 and 0.054±0.009 mM respectively). In addition, adenosine is a poor inhibitor with a $K_i$ value of 9 mM. These data strengthen the hypothesis that binding of the NADPH molecule occurs mainly through the nicotinamide ring. On the other hand, $K_i$ values for AMP and ADP-ribose, which are NAD analogs, are about 11 and 9 times smaller than the $K_i$ values for 2', 5' ADP and ATP-ribose respectively. These observations strongly suggested that the presence of the 2'-phosphate group of the NADP analogs induces a destabilization of the interaction with the Fre polypeptide. This is also consistent with a smaller $K_d$ value for NADH, which has been estimated to be lower than its $K_m$ value (8±1 µM, see above), compared to that for NADPH (54±9 µM). Furthermore, the finding that the $K_i$ value for ADP-ribose is identical to that for AMP and that the $K_i$ value for ATP-ribose is identical to that of 2', 5' ADP suggest that the phosphate ribose part of the ADP-ribose and ATP-ribose molecules does not contribute significantly to the binding. Adenosine and nicotinamide were also found to be competitive inhibitors with respect to NADPH. However they are characterized by very high $K_i$ values (9 and 20 mM respectively, Table 3).

*FAD reduction is specific for NADH*

Table 4, shows the kinetic parameters for the enzymatic reduction of FAD by NADPH, NADH or NMNH. The $k_{cat}/K_m$ values reported for the three reduced pyridine nucleotides clearly indicate that with FAD as the electron acceptor, Fre became highly NADH specific (Table 4). The kinetic parameters for the reaction (Table 4) can be compared to those for the reduction of Rf (Table 1) : when NADH was used as the electron donor, while $k_{cat}$ and $K_m$ values for the two flavins, FAD and Rf, were in the same range, the $K_m$ value for NADH was much larger in the case of FAD (301±40 μM compared to 8±1 μM in the case of Rf). With NADPH or NMNH as the electron donors this effect was much more pronounced, with $K_m$ values at around 15-20 mM (Table 4), as compared to $K_m$ values of 32±2 and 113±7 μM for NADPH and NMNH respectively with Rf as electron acceptor (Table 1). $k_{cat}$ values, in the case of NADPH and NMNH, were smaller with FAD than with Rf as the electron acceptor and thus became comparable to the $k_{cat}$ value obtained with the NADH-FAD system (Table 4). On the whole, the increased specificity for NADH compared to NADPH and NMNH, in the case of FAD reduction, resulted mainly from a much larger effect

on the $K_m$ values for NADPH and NMNH, whereas $k_{cat}$ values were only slightly affected.

*Stereospecificity of the hydride transfer*

Both stereospecifically labeled [4R-$^2$H] NADH and [4R-$^2$H] NADPH were prepared by transferring deuterium from 2-propanol-$d_8$ to NAD(P)$^+$ with appropriate NAD(P)$^+$-dependent alcohol dehydrogenases, which are specific for the transfer of the *pro-R* hydrogen of NAD(P)H. The isotopic purity of deuterated NAD(P)H was determined without purification by $^1$H-NMR spectrometry at 500 MHz. As shown in Fig. 4A, the two diastereotopic *pro-R* and *pro-S* hydrogen atoms at C-4 of the non-deuterated dihydronicotinamide ring of NADH were detected at 2.70 ppm and 2.58 ppm respectively. In the $^1$H-NMR spectrum of the deuterated analog, the signal at 2.70 ppm has disappeared almost completely indicating that deuterium has been transferred specifically to the R-position (Fig. 4B). The enzymatic deuteration was less specific in the case of NADP$^+$ and the isotopic purity of [4R-$^2$H] NADPH was found to be 90% approximately (data not shown).

The stereoselectivity of the hydrogen transfer from [4R-$^2$H] NADH to Rf catalyzed by Fre was determined from $^1$H-NMR analysis of the NAD$^+$ product, as described in Materials and Methods. The almost complete disappearance of the peaks in the 2.7-2.9 ppm region, characteristic of NADH, indicated that oxidation to NAD$^+$ was more than 95% completed (data not shown). Deuterium free NAD$^+$, which can be obtained in a control experiment with NADH, has a characteristic C-4 proton doublet at 8.74 ppm, in the aromatic region of the $^1$H NMR spectrum (Fig. 5A). The presence of this doublet, also present in the product of [4R-$^2$H] NADH oxidation, unambiguously shows that Fre preferably transfers the deuterium from the *pro-R* position resulting in the formation of deuterium free NAD$^+$, as the major product (Fig. 5B). Comparison of the intensities of the aromatic signals however showed that a minor amount (less than 10%) of deuterated NAD$^+$ was also formed. The latter could be assigned to the non-enzymatic reduction of free flavin by NADH, which was previously shown to be effective at the high concentrations of electron donor used in this study (9). Comparable results have been obtained when the stereoselectivity of the hydrogen transfer from [4R-$^2$H] NADPH to Rf catalyzed by Fre was determined under the same experimental conditions (data not shown).

These results show that the hydride transfer is *pro-R* stereospecific during reduction of Rf by NAD(P)H catalyzed by Fre.

DISCUSSION

The NAD(P)H:flavin oxidoreductase from *E.coli*, named Fre, is the prototype for the class of cofactor-free flavin reductases. The mechanism of the reaction has been partially elucidated previously with the demonstration that the flavin reductase provides a site which accommodates both reduced pyridine nucleotides and flavins, in which the hydride transfer takes place (14, 23).

In a previous paper, we identified the parameters governing the interaction of the flavin substrate with the enzyme polypeptide chain (14). The interesting conclusion was that binding of the flavin involved the isoalloxazine ring almost exclusively, with little contribution from the sugar chain. In the present work, we have studied the interaction of the pyridine nucleotide with the enzyme and show that Fre has also some unique properties with regard to that interaction.

Although Fre can operate with both NADH and NADPH when using Rf as the electron acceptor, we have found that the presence of the 2'-phosphate group had a destabilizing effect on the binding of the pyridine nucleotide to Fre. This is clear from the following observations : i) $K_d$ value for NADH has been estimated to be smaller than its $K_m$ value (8±1 µM), reflecting a better affinity of the protein for NADH than for NADPH ($K_d$ value of 54±9 µM); ii) $K_d$ values, for AMP and ADP-ribose were 10 times smaller than the $K_d$ values for 2', 5' ADP and ATP-ribose.

On the other hand, our results strongly suggested that the adenosine-ribose moiety of the NAD(P)H molecules is not important for the interaction between the pyridine nucleotide molecule and the protein. As a matter of fact, the $K_d$ values for AMP or ADP-ribose moiety of the NADH molecule and for 2', 5' ADP or ATP-ribose moiety of the NADPH molecule are much higher than the $K_d$ values estimated and determined for the corresponding substrate, NADH and NADPH. That the nicotinamide ring itself could have a marked contribution to the binding of the pyridine nucleotide to Fre was further supported by the following data.

First, the affinity of the pyridine nucleotide is strongly affected by modification of the carboxamide group of the nicotinamide ring, as illustrated by the larger differences between $K_m$ and $K_d$ values determined for 3-acetyl NADH, thio-NADH and the values estimated for NADH (Table 1). Second, NMNH, which lacks the adenosine part of the NAD(P)H molecule, was found to be a very good substrate for Fre and exhibited a $K_d$ value only 2 times higher than that of NADPH (Table 1). Taking into account that the same $K_d$ values were obtained for AMP and ADP-ribose on one side and for 2', 5' ADP and ATP-ribose on the other side, one could conclude that binding of NMNH occurs mainly by its nicotinamide ring with no marked contribution of its phosphate ribose moiety.

Taken together, all these results clearly confirmed the presence of a specific recognition site for the nicotinamide ring on the Fre polypeptide chain with significant contribution to the binding of the reduced pyridine nucleotide.

As mentioned above, the $K_d$ values for NMNH and NADPH were found to be comparable. This suggests that the negative effect of the NADPH 2'-phosphate group on the protein-substrate interaction cancels almost completely the binding effect of the adenosine part of the molecule. These observations suggest a similar mode of binding of NADPH and NMNH, with the main interaction occurring with the nicotinamide ring. NADH, in contrast, binds to Fre through interactions involving both the nicotinamide ring and the adenosine part of the molecule.

The difference between NADPH, NMNH, on one side, and NADH on the other side, with regard to binding to Fre is also reflected in the kinetic parameters determined in the presence of FAD as the electron acceptor (Table 4). The data show that FAD induces a dramatic increase of the $K_m$ values for NADPH and NMNH, whereas $K_m$ value for NADH is much less affected. Even though the basis of this difference is difficult to identify, these data further support the notion of a comparable binding mode of NADPH and NMNH.

In addition, it seems that the order of release of the products from this enzyme can be different from one pyridine nucleotide to another. With NMNH and NADPH as electron donors, the enzyme mechanism displays the same order of product release, with the reduced flavin released first followed by $NMN^+$ or

NADP$^+$. This is in contrast with NADH, for which the order of product release is reversed. However, the present data gives no clear indication which might explain these differences.

We have determined that the enzymatic hydride transfer is stereospecific for the C-4 pro-R hydrogen atom of the nicotinamide ring of the NAD(P)H and then that Fre can be classified as an A-side NAD(P)H dependent oxidoreductase. This provides further support for our suggestion that Fre is structurally and mechanistically related to the FNR family (22, 23). As a matter of fact, members of the FNR family are A-side enzymes, in contrast to members of an other flavoprotein family, the disulfide reductase family, which are B-side enzymes (35).

In the case of the flavoproteins of the FNR family, the binding mode of the pyridine nucleotide with the protein seems to be completely different from what we have found in the case of Fre. In fact, in the case of spinach FNR (36) and nitrate reductase (37), binding and inhibition studies using NAD(P) analogs, suggested that binding of the pyridine nucleotide to the proteins, involved the adenosine part of the molecule exclusively with no contribution of the nicotinamide ring. This has been partly confirmed by the structure of the spinach FNR complexed with NAD(P) analogs, which showed ordered binding of only the adenosine phosphate moiety of the pyridine nucleotide molecule (20). The crystal structure of the PDR-NADH complex, where the entire pyridine nucleotide molecule is seen, has revealed that the nicotinamide ring is not correctly positioned for hydride transfer with the flavin cofactor (38). A conformational change was postulated in order to allow direct contact of the nicotinamide ring and the flavin (38). This is supported by rapid kinetic studies on PDR which suggested a two-step mechanism for NADH binding (25, 26).

In the case of alcohol dehydrogenase from horse liver, a two-step binding mode of pyridine nucleotide has also been proposed. The first binding step would consist of docking of the adenosine part of the pyridine nucleotide and the second of conformational isomerization for the binding of the nicotinamide part (39). The fact that the nicotinamide ring is apparently not directly recognized by the active site of these NAD(P) dependent dehydrogenase is in keeping with the fact that NMN(H) has never been reported, to our knowledge, to be an efficient substrate in an enzyme NAD(P)-dependent reaction.

Recognition of the pyridine nucleotide by Fre, by both the adenosine and nicotinamide moiety (for the case of NADH) or almost exclusively by the nicotinamide moiety (for the case of NADPH) appears to be unusual. It could thus represent a new type of pyridine nucleotide binding mechanism, which has never been described previously. That this NAD(P)H binding mechanism is different from what it was suggested for the flavoproteins of the FNR family, may contribute to make Fre a flavin reductase rather than a flavoprotein, in spite of the structural and functional homologies between these two classes of protein (22, 23).

Finally, this work and a previous one (14) establishes that Fre exhibits a very unique mode of binding for both substrates which sets it apart from other NAD(P)H- and flavin-dependent enzymes. It uses a remarkable symmetrical strategy for recognition of the NAD(P)H and the flavin. In both cases, it is mainly the redox part of the substrate molecules, the nicotinamide ring and the isoalloxazine ring, which are recognized by the polypeptide chain.

ACKNOWLEDGMENTS

We are grateful to Dr. M. C. Brochier and Dr. C. Fontaine for technical assistance in $^1$H-NMR experiments and to Dr. Elspeth Gordon for reading the manuscript.

FOOTNOTES

[1] The abbreviations used are : Fre, NAD(P)H : flavin oxidoreductase; FNR, ferredoxin-NADP$^+$ reductase; PDR, phthalate dioxygenase reductase; Rf, riboflavin; AMP, adenosine 5'-monophosphate; 2', 5' ADP, 2'-phosphoadenosine 5'-phosphate; ADP-Ribose, adenosine 5'-diphosphoribose; ATP-ribose, 2'-phosphoadenosine 5'-diphosphoribose; NMNH, β-nicotinamide mononucleotide reduced form; 3-acetyl NADH, 3-acetylpyridine adenine dinucleotide reduced form; thio-NADH, thionicotinamide adenine dinucleotide reduced form; $v_i$ , initial velocity; $v_m$ , maximal velocity; $K_{mapp}$ , apparent Michaelis constant; e, enzyme concentration; $\phi_o$, $\phi_A$, $\phi_B$, $\phi_{AB}$, kinetic parameters defined by Dalziel (30); nd, not determined.

FIGURE LEGENDS

Scheme 1. Structure of the different pyridine nucleotide analogs.

Figure 1. A, flavin reductase initial velocity as a function of NMNH concentration in the presence of 1.00 (○), 1.25 (□), 2.00 (●) or 3.00 μM (△) Rf. The flavin reductase concentration, e, used in the assay was 0.0192 μM. The results are presented as double reciprocal plots with straight line determined by a linear regression program. B, secondary plot of the slopes from A against 1/[Rf] and fitted with a straight line, corresponding to the equation : $y = \phi_{AB} x + \phi_A$. C, secondary plot of the intercepts from A against 1/[Rf] and fitted with a straight line, corresponding to the equation : $y = \phi_B x + \phi_0$.

Figure 2. Double reciprocal plots of the flavin reductase initial velocity as a function of NADH concentration in the presence of 0.25 (△), 0.50 (●), 1.00 (□) or 20.00 μM (○) Rf.

Figure 3. A, $NAD^+$ as a mixed-type inhibitor for NADH. The enzyme initial velocity was assayed as a function of NADH concentrations using 1 μM Rf, in the absence (△) or in the presence of 0.47 (□), 0.95 (○) or 1.43 (●) mM $NAD^+$. B, $NAD^+$ as a mixed-type inhibitor for Rf. The enzyme initial velocity

was assayed as a function of Rf concentrations using 10 μM NADH, in the absence (△) or in the presence of 3 (○), 4 (□) or 5 (●) mM NAD$^+$.

Figure 4. $^1$H-NMR spectra (500 MHz) of NADH (A) and [4R-$^2$H] NADH (B) obtained by enzymatic reduction.

Figure 5. Aromatic region of the $^1$H-NMR spectra (500 MHz) of NAD$^+$ obtained from the Fre-catalyzed reduced pyridine nucleotide oxidation in the presence of Rf. A : NAD$^+$ obtained from NADH enzymatic oxidation. B : NAD$^+$ obtained from [4R-$^2$H] NADH enzymatic oxidation.

TABLES